\documentclass[5p,times]{elsarticle}
\usepackage{amssymb}
\usepackage{amsthm}
\usepackage{makecell}
\usepackage{amsmath}
\usepackage{color,xcolor}
\newtheorem{remark}{Remark}
\usepackage[utf8]{inputenc}
\usepackage{booktabs, caption, makecell}

\usepackage{threeparttable}
\usepackage{epigraph}
\usepackage{enumitem}
\usepackage{subfigure}
\usepackage{url}
\usepackage{color}
\usepackage{bbm}
\usepackage{amsmath, amssymb, amsfonts, amsthm}
\usepackage{siunitx}
\usepackage[linesnumbered, lined, boxed]{algorithm2e}
\newtheorem{definition}{Definition}

\usepackage{multirow}
\usepackage[figuresright]{rotating}
\usepackage{amsmath}

\usepackage{tablefootnote}

\newtheorem*{Lyapunov search problem}{Lyapunov function search problem}

\begin{document}

\begin{frontmatter}

\title{A constrained symbolic regression approach for Lyapunov function discovery}
\author[inst1]{Ilias Mitrai\corref{cor1}}
\ead{imitrai@che.utexas.edu}

\author[inst2]{Wentao Tang}

\cortext[cor1]{Corresponding author}

\affiliation[inst1]{%
  organization={McKetta Department of Chemical Engineering, The University of Texas at Austin},
  city={Austin},
  postcode={TX 78712},
  country={United States}
}

\affiliation[inst2]{%
  organization={Department of Chemical and Biomolecular Engineering, North Carolina State University},
  city={Raleigh},
  postcode={NC 27695},
  country={United States}
}

\begin{abstract} 
In this paper, we consider the data-driven discovery of Lyapunov functions for autonomous dynamical systems. We represent the Lyapunov function as an expression tree of fixed depth and formulate the Lyapunov discovery task as a constrained self-supervised symbolic regression problem. The constraints model the output of the Lyapunov function for a given input as well as the Lyapunov stability conditions. This modeling approach makes no a priori assumptions about the functional form of the Lyapunov function, is inherently interpretable since the function is obtained in a symbolic form, and, in principle, can be applied to any continuous dynamical system. We also develop a tailored branch-and-bound-and-check solution approach to efficiently solve the resulting learning task. Applications to several case studies show the ability of the proposed approach to discover Lyapunov functions. 
\end{abstract}

\end{frontmatter}

\section{Introductions} \label{sec:intro}

Lyapunov function is a fundamental concept of control theory and a necessary tool for analyzing the stability of dynamical systems \cite{khalil2002nonlinear}, designing controllers \cite{sontag1989universal} (including model predictive controllers \cite{grimm2005model, heidarinejad2012economic}), and safe reinforcement learning algorithms \cite{chow2018lyapunov}. 
Lyapunov-like inequalities naturally also play a key role in the verification of passivity, dissipativity, finite-gain stability, input-to-state stability,  contraction, and safety properties \cite{sontag1998mathematical, haddad2008nonlinear, ames2019control, tsukamoto2021contraction}. 
In particular, for a dynamical system, the existence of a Lyapunov function provides a certificate of stability of the equilibrium, as presented below.
\begin{definition}[Lyapunov function \cite{khalil2002nonlinear}]
Let $x=0$ be an equilibrium point for $\dot{x}=f(x)$, with $f: \mathcal{D} \to \mathbb{R}^{n}$, and $V: \mathcal{D}\to \mathbb{R}$ be a continuously differentiable function on a neighborhood $\mathcal{D}$ of $x=0$ such that
\begin{equation}\label{eq:lyap-psd}
    V(0)=0, \ V(x)>0 \ \forall x \in \mathcal{D}\setminus \{0\}
\end{equation}
and
\begin{equation}\label{eq:lyap-deriv}
    \dot V(x)=\nabla V(x)^\top f(x) < 0 \ \forall x \in \mathcal{D} \setminus  \{0\},
\end{equation}
then $x=0$ is asymptotically stable. 
\end{definition}
Despite the widespread use and importance of Lyapunov functions, their discovery is challenging. Traditionally, Lyapunov functions have been discovered using domain knowledge and intuition \cite{gurel1969guide}. Although this trial-and-error approach can discover Lyapunov functions, it is time-consuming. This has motivated the development of automated approaches for constructing/discovering Lyapunov functions \cite{giesl2015review}. The search for a Lyapunov function can be posed as the following search problem.
\begin{Lyapunov search problem}
    \normalfont Given a dynamical system described by an ordinary differential equation $\dot{x}(t)=f(x(t))$, find $V(x)$ such that
    \begin{equation*}
        \begin{aligned}
            & \dot{V}(x) = \nabla_{x} V(x)^\top f(x) <  0, \ \ \forall x \in \mathcal{D}\setminus \{0\}\\
            & V(x) >0, \ \ \forall x \in \mathcal{D}\setminus \{0\}\\
            & V(0)=0.
        \end{aligned}
    \end{equation*}
\end{Lyapunov search problem}
The solution of this problem presents two challenges. First, it has an infinite number of constraints, since the Lyapunov conditions must be satisfied at all points in the domain of the state variables. The second challenge is related to the representation of the Lyapunov function, i.e., the search variable in this problem is a function. In other words, the problem of interest is to find \emph{functions} that satisfy \emph{functional constraints} \cite{tang2019bilevel, tang2023optimal}. 

Several automated computational approaches have been proposed to search for a Lyapunov function \cite{giesl2015review}. For example, for a linear time invariant (LTI) system, $\dot{x}=Ax$, a candidate Lyapunov function is $V(x)=x^\top P x$, where the matrix can be computed by solving the Lyapunov equation. For polynomial dynamical systems, the standard approach is to use sum-of-squares optimization \cite{papachristodoulou2002construction, goubault2014finding}. In these cases, the semi-infinite nature of the problem is addressed by imposing algebraic constraints, such as $P\succeq 0$ for LTI systems or by replacing the Lyapunov conditions with sum-of-squares (SOS) constraints in the polynomial case. 
Despite these advances, the discovery of a Lyapunov function for a general nonlinear system remains challenging, since its functional form is not known a priori. An approach is to parameterize the Lyapunov function using basis functions, such as piece-wise affine \cite{julian1999parametrization, marinosson2002lyapunov, bjornsson2014computation}, global basis functions \cite{ johansen2000computation, giesl2008construction}, Bernstein polynomials \cite{ben2016linear}, and general polynomials \cite{ratschan2010providing,li2026learning}, leading to linear, convex, and mixed integer linear programming formulations. 
An alternative approach is to parameterize the Lyapunov function using neural networks, which are universal function approximators \cite{petridis2006construction, serpen2005empirical, richards2018lyapunov, chang2019neural, chen2021learning, zhou2022neural, wu2023neural, grande2023augmented, abate2020formal, chen2020learning}. In such cases, the learning task is reformulated as an unconstrained problem by penalizing the violation of the Lyapunov conditions. Although such neural network approaches have been applied to various problems, the resulting function is a black box. 

A challenge with most basis-function and neural-network approaches is certifying the validity of the learned Lyapunov function, i.e., ensuring that the candidate function satisfies the Lyapunov conditions across the entire domain. This challenge arises from enforcing or promoting the satisfaction of the Lyapunov conditions only at specific points of the domain, i.e., the available data. 
In basis-function-like approaches that leverage constrained optimization, the Lyapunov conditions are enforced only on the sampled trajectories, whereas for the neural network-based approaches, the returned function is not guaranteed to satisfy the Lyapunov conditions even for the given data, since the violation of the conditions is penalized in the objective. This has motivated the development of iterative loops in which, after the search problem terminates, a verification step checks whether the identified function is a valid Lyapunov function \cite{dai2020counter}. This step can be computationally expensive for the neural network-based approaches since a neural network verification problem must be solved. 

The validity issue can be partially relieved by \emph{kernel methods} used extensively in the second author's recent works \cite{tang-ye2025koopman, tang-ye2026koopman, tang-ye2026dissipativity}. Therein, a reproducing kernel Hilbert space (RKHS) is defined to include all state-dependent functions that have zero values at the origin and a global smoothness property, and hence the Lyapunov function is postulated as a sum-of-squares form on that RKHS and thus represented by a nonnegative operator. 
Under a given required decay rate, the nonnegative operator can be found under the estimated Koopman operator (i.e., a linear operator on the RKHS capturing the nonlinear dynamics on finite-dimensional $\mathcal{D}$) \cite{tang-ye2026koopman} or directly based on the snapshot data \cite{tang-ye2026dissipativity}. 
Due to the smoothness properties of such postulated Lyapunov functions as ``kernel SOS'', the validity on the data points, when the sample fills the space densely enough, implies an approximate validity everywhere \cite{tang-ye2026dissipativity}. 

\begin{figure}[!t]
    \centering
    \includegraphics[scale=0.8]{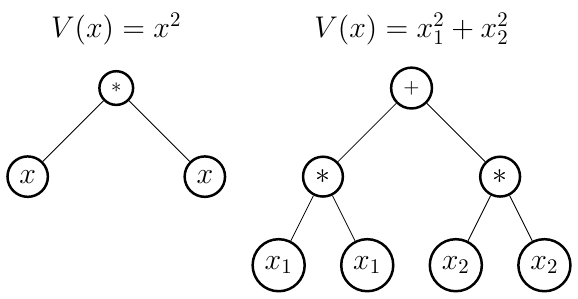}
    \caption{Symbolic expression tree representation of two Lyapunov functions.}
    \label{fig:Lyapunov examples}
\end{figure}
An alternative is to model the Lyapunov function as an expression tree as presented in Fig.~\ref{fig:Lyapunov examples}. This representation is inherently interpretable since the Lyapunov function can be obtained in symbolic form, and can represent any function, given that the depth of the tree and the mathematical operations considered are adequate \cite{knuth1997art,smith1996global,smith1999symbolic}. Unlike basis function and neural network-based approaches, which search over the parameter space of each candidate model, the search over the space of all possible functions, i.e., expression trees, is combinatorial. Traditionally, symbolic regression problems were solved using genetic programming by evolving an original set of expressions randomly for a given number of iterations \cite{koza1994genetic, grosman2006lyapunov, mcgough2010symbolic}. This search approach is inefficient since it is random and cannot guarantee that the returned function is a valid Lyapunov function for the given data. Recently, mixed-integer optimization approaches have been proposed to solve the symbolic regression task \cite{austel2017globally, cozad2018global, neumann2020new, kim2023learning}.

In this paper, we pose the Lyapunov discovery task as a \emph{constrained self-supervised symbolic regression problem}. This approach enables the systematic modeling of all candidate functions for a given set of operators and tree depth, and the systematic search over the discrete function space using branch-and-bound techniques. Classical symbolic regression is a supervised learning problem, since one seeks a function that minimizes the prediction error between the predicted and true label. However, the data-driven discovery of a Lyapunov function is a self-supervised learning problem, since the output of the Lyapunov function is not measured; i.e., the available data contains only information about the evolution of the states. Therefore, the identification of a valid function relies on the presence of constraints. In the proposed approach, these constraints are explicitly expressed as functions of the mathematical operators assigned in the expression tree.

The contributions of this paper are as follows.
\begin{itemize}
    \item We model the Lyapunov function as an expression tree of fixed depth. This modeling approach can represent many functional forms of the Lyapunov function using simple mathematical operations such as addition, subtraction, division, and multiplication.
    \item We enforce the Lyapunov conditions for the given data, i.e., the output of the Lyapunov function is always non-negative and its derivative is negative semi-definite, as constraints in the learning task. The resulting learning problem is a mixed-integer nonlinear programming problem, which can be solved with state-of-the-art branch-and-bound-based global optimization algorithms. The objective can be either the complexity of the returned function, i.e., find the simplest Lyapunov function for the given data or the convergence rate.
    \item We develop a hybrid branch-and-bound-and-check algorithm to efficiently solve the learning problem. Unlike existing approaches which certify the candidate function after the learning task is solved, we embed the verification step inside the search. This is possible due to the symbolic tree representation of the tree and the nature of branch-and-bound solvers, which find integer feasible solutions, i.e., candidate Lyapunov functions, during the solution process. Specifically, during branch and bound, once an integer feasible solution is found at a node of the branch and bound tree, we check if the function is a valid Lyapunov function for the entire domain and stop if the required conditions are satisfied, i.e., $V(x) \geq 0$ and $dV(x)/dt \leq 0$; otherwise, the branch-and-bound search continues.
\end{itemize}

We note that recently, several approaches have been proposed to identify Lyapunov functions symbolically using transformers \cite{alfarano2024global}, reinforcement learning for building the expression tree \cite{zou2025analytical}, and distilling interpretable equations from neural Lyapunov functions \cite{feng2024combining}. Although these approaches lead to interpretable Lyapunov functions, they are local search approaches since they do not solve the Lyapunov function discovery task to global optimality. 

The rest of the document is organized as follows: In Section~\ref{sec: proposed learning approach} we present the proposed symbolic tree representation of the Lyapunov function and the MINLP learning task, in Section~\ref{sec: branch-and-bound-and-check solution} we present the proposed branch-and-bound-and-check solution approach, and in Section~\ref{sec: computational resulrs} we use the proposed approach to discover Lyapunov functions for different case studies.

\section{Symbolic discovery of Lyapunov functions}\label{sec: proposed learning approach}
We consider a autonomous dynamical system $\dot{\textbf{x}}=f(\textbf{x})$, with $\textbf{x} \in \mathbb{R}^{\rm{N_{x}}}$, $\rm{N_{x}}$ is the number of states, and $f: \mathbb{R}^{\rm{N_{x}}}\mapsto \mathbb{R}^{\rm{N_{x}}}$ is a smooth vector function. We assume that $\rm{N_{t}}$ trajectories of the system are available, denoted as $\mathcal{I}$, from multiple initial conditions $\textbf{x}_{0}$. For each trajectory $i \in \mathcal{I}$, the states are sampled at $\rm{N_{i}}$ points. We define as $\mathcal{J}$ the set of data points in each trajectory, $x_{kij}$ as the value of state $k$ for initial condition $i$ at data point $j$, and $\mathcal{J}_{i}$ as the set of data points in the trajectory starting from initial condition $i$ for which $|\textbf{x}_{ij}|= 0$ ($\mathcal{J}_{i} \subseteq \mathcal{J}$). Overall, the training data are $\{\textbf{x}_{ij}\}_{i=1, j=1}^{N_{t}, N_{i}}$.

\subsection{Sets and variables}
\begin{figure}[h]
    \centering
    \includegraphics[scale=0.8]{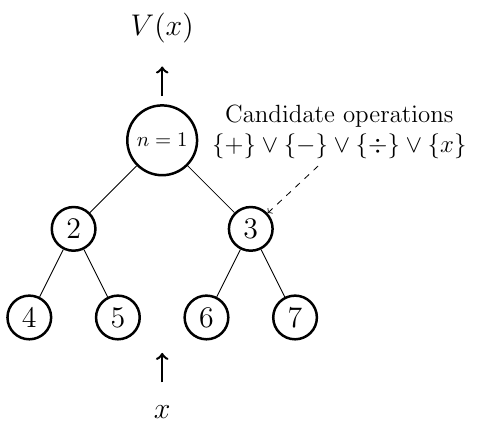}
    \caption{Symbolic expression tree representation of the Lyapunov function and candidate operations at each node.}
    \label{fig:Lyapunov idea}
\end{figure}

The Lyapunov function is modeled as a perfect binary tree of depth $d$ as presented in Fig.~\ref{fig:Lyapunov idea}.

We define as $\mathcal{N} = \{1,\dots, 2^{d+1}-1\}$ the set of nodes in the tree, $\mathcal{T}=\{2^{d},\dots,2^{d+1}-1\}$ as the set of terminal (leaf) nodes, and $\mathcal{N}_{\rm{T}} = \mathcal{N}\setminus \mathcal{T}$ as the set of non-leaf nodes. In this representation, each node has two children, $2n$ and $2n+1$, except the leaf nodes $\mathcal{T}$. At each node, a binary $\mathcal{B} = \{+,-,*,/\}$, unary $\mathcal{U} = \{ \exp, \log\}$ operator, or operand $\mathcal{L} = \{x_1,...x_{N_x}, \rm{cst}\}$ can be assigned ($\rm{cst}$ represents the constant operator). We define as $\mathcal{O} = \mathcal{B} \cup \mathcal{U} \cup \mathcal{L}$ the set of all possible operators and operands that can be assigned to a node. 

We define a binary variable $y_{on}$ which is equal to one if operator $o$ is assigned to node $n$ and zero otherwise. We also define $c_{n} \in [c_{\rm{lb}}, c_{\rm{ub}}]$ as the value of the constant for node $n$. These variables exactly parameterize the Lyapunov function, since once fixed, determine the functional form and the values of the constants. Finally, we define $v_{ijn} \in [v_{\rm{lb}}, v_{\rm{ub}}]$ as the value at node $n$ for initial condition (trajectory) $i$ and data point $j$. Under this parameterization, the output of the Lyapunov function for data point $j$ in trajectory $i$ is equal to the value of the root node $v_{ij1}$.

\subsection{Tree defining constraints}
The problem formulation has three sets of constraints. The first, called tree-defining, guarantees that the assignment of operators to the nodes satisfies arity rules. These constraints are adapted from \cite{cozad2018global,kim2023learning}. First, we enforce that at each node $n$ at most one operator can be assigned via the following constraints
\begin{equation}\label{eq: one operator}
    \sum_{o \in \mathcal{O}} y_{on} \leq 1 \ \ \forall n \in \mathcal{N}.
\end{equation}
The assignment of an operator at a node depends on the operators of the parent and children nodes. For example, a binary or unary operator can be assigned only to non-leaf nodes $\mathcal{N}_{T}$. These requirements are enforced via the following constraints
\begin{equation} \label{eq: operator assignment 1}
    \begin{aligned}
        \sum_{o \in \mathcal{B} \cup \mathcal{U}} y_{on} & = \sum_{o \in \mathcal{O}} y_{o,2n+1} \ \ \forall n \in \mathcal{N}_{T}
    \end{aligned}
\end{equation}
\begin{equation} \label{eq: operator assignment 2}
    \begin{aligned}
        \sum_{o \in \mathcal{B}} y_{on} & = \sum_{o \in \mathcal{O}} y_{o,2n} \ \ \forall n \in \mathcal{N}_{T}.
    \end{aligned}
\end{equation}
The first constraint (Eq.~\ref{eq: operator assignment 1}) guarantees that if a binary or unary operator is assigned to node $n$, then one operator is assigned to node $2n+1$. The second constraint enforces (Eq.~\ref{eq: operator assignment 2}) that if a binary operator is assigned to a node $n$ then an operator or operand is assigned to node $2n$. 

Finally, we enforce that all the state variables must be present in the final expression of the Lyapunov function via the following constraints
\begin{equation} \label{eq: assign x variables}
    \sum_{n \in \mathcal{N}} y_{on} \geq 1 \ \ \forall o \in \mathcal{L}\setminus \{\rm{cst}\}.
\end{equation}

\subsection{Value defining constraints}
The second set of constraints, called value-defining, guarantees that the output of the tree is computed correctly based on the assignment of the operators. For example, if addition is assigned at the root node, $y_{+,1}=1$, then the value of the node for data point $j$ in trajectory $i$ is equal to the summation of the values of the children nodes, i.e., $v_{ij1}=v_{ij2}+v_{ij3}$. However, the assignment of the operators to the nodes is a degree of freedom; thus, the value-defining constraints for the addition operator are enforced via the following big-M constraints
\begin{equation} \label{eq: value def addition}
    \begin{aligned}
        v_{ijn} - (v_{ij,2n} + v_{ij,2n+1}) & \leq \overline{M}_{+} (1-y_{+,n}) \ \ \forall i \in \mathcal{I}, j \in \mathcal{J}, n \in \mathcal{N}_{T}\\
        v_{ijn} - (v_{ij,2n} + v_{ij,2n+1}) & \geq \underline{M}_{+} (1-y_{+,n}) \ \ \forall i \in \mathcal{I}, j \in \mathcal{J}, n \in \mathcal{N}_{T}.
    \end{aligned}
\end{equation}
Similarly, we can write the constraints for the constant (Eq.~\ref{eq: value defining cst}), subtraction (Eq.~\ref{eq: value defining sub}), multiplication (Eq.~\ref{eq: value defining mult}), and division (Eq.~\ref{eq: value defining division}) operators as presented below:
\begin{table}[]
\centering
\caption{Upper and lower bounds for the big-M value defining constraints \cite{cozad2018global}}
\label{table: big-M}
\begin{tabular}{cll}
\hline 
Operator   & $\overline{M}$ & $\underline{M}$ \\ \hline
 $+$       &  $v_{\rm{ub}}-2v_{\rm{lb}}$ &    $v_{\rm{lb}}-2v_{\rm{ub}}$  \\
 $-$       &  $2v_{\rm{ub}}-v_{\rm{lb}}$   &  $2v_{\rm{lb}}-v_{\rm{ub}}$               \\
 $*$       &  $v_{\rm{ub}}- \min\{v_{\rm{lb}}^2, v_{\rm{lb}}v_{\rm{ub}}, v_{\rm{ub}}^{2}\}$ &   $v_{\rm{lb}}- \max\{v_{\rm{lb}}^2, v_{\rm{ub}}^{2}\}$                              \\
  $/$      & $\max\{v_{\rm{lb}}^2, v_{\rm{ub}}^{2}\}$   &  $\min\{v_{\rm{lb}}^2, v_{\rm{lb}}v_{\rm{ub}}, v_{\rm{ub}}^{2}\}$ \\
$\rm{cst}$ & $v_{\rm{up}} - c_{\rm{lb}}$        &   $v_{\rm{lb}} - c_{\rm{ub}}$      \\
   \hline
\end{tabular}
\end{table}
\begin{equation}\label{eq: value defining cst}
\begin{aligned}
        v_{ijn} - c_{n} & \leq \overline{M}_{\rm{cst}}(1-y_{cst,n}) \ \ \forall i \in \mathcal{I}, j \in \mathcal{J}, n \in \mathcal{N}\\
        v_{ijn} - c_{n} & \geq \underline{M}_{\rm{cst}}(1-y_{cst,n}) \ \ \forall i \in \mathcal{I},j \in \mathcal{J}, n \in \mathcal{N}
    \end{aligned}
\end{equation}
\begin{equation}\label{eq: value defining sub}
   \begin{aligned}
        v_{ijn} - (v_{ij,2n} - v_{ij,2n+1}) & \leq \overline{M}_{-}(1-y_{-,n}) \ \ \forall i \in \mathcal{I},j \in \mathcal{J}, n \in \mathcal{N}_{T}\\
        v_{ijn} - (v_{ij,2n} - v_{ij,2n+1}) & \geq \underline{M}_{-}(1-y_{-,n}) \ \ \forall i \in \mathcal{I},j \in \mathcal{J}, n \in \mathcal{N}_{T}
    \end{aligned}
\end{equation}
\begin{equation}\label{eq: value defining mult}
\begin{aligned}
        v_{ijn} - (v_{ij,2n} v_{ij,2n+1}) & \leq \overline{M}_{*}(1-y_{*,n}) \ \ \forall i \in \mathcal{I},j \in \mathcal{J}, n \in \mathcal{N}_{T}\\
        v_{ijn} - (v_{ij,2n} v_{ij,2n+1}) & \geq \underline{M}_{*}(1-y_{*,n}) \ \ \forall i \in \mathcal{I},j \in \mathcal{J}, n \in \mathcal{N}_{T}
    \end{aligned}
\end{equation}
\begin{equation} \label{eq: value defining division}
\begin{aligned}
    & v_{ijn} v_{ij,2n+1} - v_{ij,2n} \leq  \overline{M}_{/}(1-y_{/,n}), \ \  i \in \mathcal{I}, j \in \mathcal{J},n \in \mathcal{N}_T\\
    & v_{ijn} v_{ij,2n+1} - v_{ij,2n} \geq \underline{M}_{/} (1-y_{/,n}), \ \  i \in \mathcal{I}, j \in \mathcal{J},n \in \mathcal{N}_{T}\\
    & \epsilon y_{/,n} \leq v_{ij,2n}^{2}, \ \  i \in \mathcal{I},j \in \mathcal{J}, n \in \mathcal{N}_T\\
    & \epsilon y_{/,n} \leq v_{ij,2n+1}^{2}, \ \  i \in \mathcal{I},j \in \mathcal{J}, n \in \mathcal{N}_{T}.
\end{aligned}
\end{equation}
The big-M constants for each operator are presented in Table~\ref{table: big-M} and obtained from \cite{cozad2018global}. Finally, we include constraints that ensure that if a variable $x$ is assigned to a node $n$, i.e., $y_{x,n}=1$, then for all data points, the value of that node equals the value of the variable for each data point. This is enforced via the following constraints
\begin{equation} \label{eq: value defining x 1}
\begin{aligned}
    v_{ijn} &  \leq \sum_{k \in \mathcal{L}\setminus \{\rm{cst}\}} x_{kij} y_{x_{k},n} + v_{\rm{ub}} \sum_{o \in \overline{\mathcal{O}}} y_{no} \ \ \forall i \in \mathcal{I}, j \in \mathcal{J}_{i}, n\in \mathcal{N}
\end{aligned}
\end{equation}  
\begin{equation} \label{eq: value defining x 2}
\begin{aligned}
    v_{ijn} & \geq \sum_{k \in \mathcal{L}\setminus \{\rm{cst}\}} x_{kij} y_{x_{k},n} + v_{\rm{lb}} \sum_{o \in \overline{\mathcal{O}}} y_{no} \ \ \forall i \in \mathcal{I}, j \in \mathcal{J}, n\in \mathcal{N},
\end{aligned}
\end{equation}  
where $\Tilde{\mathcal{O}} = \mathcal{B} \cup \mathcal{U} \cup \{\rm{cst}\}$. If a variable $x_{k}$ is not assigned at a node, then these constraints reduce to lower and upper bounds for the $v_{ijn}$ variables.

\subsection{Lyapunov-defining constraints} \label{sec: lyapunov defining constraints}
The constraints presented so far guarantee that the assignment of the operators is correct and that the output of the tree is computed correctly. The last set of constraints guarantees that the assignment of the operators leads to a valid Lyapunov function for the given data. First, we enforce that the output of the Lyapunov function, $v_{ij1}$, is positive for data points for which $|\textbf{x}|\neq 0$ and zero otherwise. This is presented in the following constraints
\begin{equation} \label{eq: K class function}
\begin{aligned}
    v_{ij1} & \geq \|\textbf{x}_{ij}\|^{2}  \ \ \forall i \in \mathcal{I}, j \in \mathcal{J} \setminus \mathcal{J}_{i}\\
    v_{ij1} & = 0 \ \ \forall i \in \mathcal{I}, j \in  \mathcal{J}_{i}.
\end{aligned}
\end{equation}
\begin{remark}
    \normalfont We note that if the trajectories are not long enough and the equilibrium has not been achieved, an auxiliary data point can be added which corresponds to $\textbf{x}=\textbf{0}$, assuming that the equilibrium is $\textbf{x}=\textbf{0}$. The incorporation of this data point is needed to satisfy the Lyapunov condition $V(\textbf{0})=0$.
\end{remark}

The last set of constraints (Eq.~\ref{eq: neg def V}) enforce that the derivative of the Lyapunov function is negative semi-definite. Although in principle, one can model the derivative of the function using a second expression tree \cite{engle2022deterministic}, this approach increases the complexity of the learning problem. To improve the tractability, we approximate the time derivative of the Lyapunov function using finite differences, as presented below
\begin{equation}\label{eq: neg def V}
\begin{aligned}
    \frac{v_{i,j+1,1} - v_{ij1}}{\Delta t} & \leq -\alpha \|x_{ij}\|^{2} \ \ \forall i\in\mathcal{I}, j \in \mathcal{J}\setminus\mathcal{J}_{i}\\
    v_{i,j+1,1} - v_{ij1} & = 0 \ \ \forall i\in\mathcal{I}, j \in \mathcal{J}_{i},
\end{aligned}
\end{equation}
where $\alpha$ is the rate of convergence and $\|x_{ij}\|^{2}$ is a ``$\mathcal{K}$-class'' function of $x$. This condition guarantees asymptotic stability for the equilibrium. Depending on the system, different conditions can be used and expressed as constraints; e.g., exponential stability would lead to the following constraint:
\begin{equation*}
    \frac{v_{i,j+1,1} - v_{ij1}}{\Delta t}  \leq -\alpha v_{ij1} \ \ \forall i\in\mathcal{I}, j \in \mathcal{J} \setminus \mathcal{J}_{i}.
\end{equation*}

\subsection{Overall learning problem}
Given these constraints, different learning problems can be formulated depending on the desired outcome. For example, if the goal is to find the Lyapunov function with the maximum convergence rate, then the objective to be minimized is $-\alpha$. Alternative objective functions can include the complexity of the Lyapunov function, captured by the number of operations that must be performed to compute the output, and/or a weighted combination as presented below
\begin{equation} \label{eq: learning task}
    \begin{aligned}
    \min_{y_{on}, c_{n}, v_{ijn}, 
    \alpha} \ \ & -w_{1} \alpha + w_{2} \sum_{n \in \mathcal{N}} \sum_{o \in \mathcal{O}} y_{no} + w_{3} \sum_{n \in \mathcal{N}} c_{n}^{2}\\
    \text{s.t.} \ \ & \mathrm{Eq.}~\ref{eq: one operator}, \ref{eq: operator assignment 1}, \ref{eq: operator assignment 2}, \ref{eq: assign x variables}, \ref{eq: value def addition}, \ref{eq: value defining cst}, \ref{eq: value defining sub}, \ref{eq: value defining mult}, \ref{eq: value defining division}, \ref{eq: value defining x 1}, \ref{eq: value defining x 2}, \ref{eq: K class function}, \ref{eq: neg def V}\\
    & y_{on}\in\{0,1\}, v_{ijn} \in [v_{\rm{lb}}, v_{\rm{ub}}], c_{n} \in [c_{\rm{lb}}, c_{\rm{ub}}], \alpha \geq 0,
    \end{aligned}
\end{equation}
with $w_{1}$, $w_{2}$, $w_{3}$ being weights. 

\begin{remark}
    \normalfont The proposed learning formulation assumes that the data are noise-free, i.e., the value of the Lyapunov function decreases monotonically over time. If one had only access to experimental data, then the proposed formulation should be altered by using slack variables in Eq.~\ref{eq: neg def V} to allow $V(\textbf{x})$ to increase temporarily due to the noise. That is, $v_{i,j+1,1}\leq v_{i,j,1}-\alpha\|x_{ij}\|^2 + \varsigma$ for a small $\varsigma$. These slack variables should also be penalized in the objective function with an appropriate weight factor. 
\end{remark}

\begin{remark}
    \normalfont The Lyapunov function obtained by solving the learning task presented in Eq.~\ref{eq: learning task} is not guaranteed to be valid for the given data or the entire domain. Regarding the former, the inability to guarantee satisfaction of the Lyapunov conditions stems from approximating $dV/dt$ using finite differences. Regarding the entire domain, the proposed approach cannot guarantee a valid Lyapunov function even if $dV/dt$ is computed exactly, due to the limited number of data points (trajectories). We note that these limitations are present in almost all data-driven Lyapunov discovery approaches and can be overcome by using a verification step. 
\end{remark}

\begin{remark}
    \normalfont In the proposed approach, it is assumed that the model is known, i.e., the differential equations governing the dynamic behavior of the system. This assumption is necessary to certify that the candidate function returned is a valid Lyapunov function for the entire domain. If the model is unknown then the proposed approach can still be used to search for a Lyapunov function, but the verification step cannot be applied to certify the candidate function for the entire domain.
\end{remark}

\section{Branch-and-bound-and-check solution approach}\label{sec: branch-and-bound-and-check solution}

The learning problem presented above is a mixed-integer nonlinear programming problem whose complexity depends on the depth of the expression tree, the number of state variables, the types of operators used, and the number of data points. As mentioned above, although the expression tree parameterization of the Lyapunov function is exact, the solution of the learning problem is not guaranteed to return a valid Lyapunov function for the entire domain due to the finite number of data points and the finite-difference approximation of $dV/dt$. An approach to overcome this limitation is to use a verification step after the problem is solved, which is common in learning-based approaches for Lyapunov function discovery. However, this approach can be computationally expensive, as solving the MINLP to global optimality can be time-consuming. 

To improve the computational performance of the search problem, we incorporate the verification step inside the branch-and-bound search approach. The main idea is to combine branch-and-bound with \emph{branch-and-check}. Specifically, we start by solving the learning task (optimization problem) using standard spatial branch-and-bound (see Algorithm~\ref{algo: branch and bound and check}). The integer feasible solutions found during branch-and-bound correspond to candidate Lyapunov functions which follow arity rules, compute $V(\textbf{x})$ correctly, they are positive definite $V(\textbf{x})>0$, $V(0)=0$, and satisfy the approximate equations for $dV/dt$ for the given data. However, this candidate function is not guaranteed to be valid for the entire domain. 

To check the validity of the proposed function, we add the following steps in the branch-and-bound procedure: Once a integer feasible solution is found at a node $p$, the validity of the candidate function for the entire domain is assessed by checking if the minimum of $V(x)$ is positive, i.e., $\min_{x \in \mathcal{D}} V(x) \geq 0$, if the maximum of $\dot{V}(x)$ is negative, i.e., $\max_{x \in \mathcal{D}} \dot{V}(x) \leq 0$, and if $V(0)=0$. If these conditions are satisfied, then the function is a Lyapunov function for the entire domain. At this point, one can either stop the search, since a valid Lyapunov function is found or continue branch and bound until a new Lyapunov function with a better objective function value is found. If, at node $p$, the candidate function is not a valid Lyapunov function over the entire domain, the branch-and-bound procedure continues. 

\begin{algorithm}[t]
\SetAlgoLined
\KwIn{Optimization problem}
\KwOut{Lyapunov function $V(x)$}
Solve the continuous relaxation of the problem\;
Obtain set of open nodes $R$\;
\While{{function not found}}{
Select a node $\rho$ from $R$ using node selection techniques\;
Solve continuous relaxation at $\rho$\;
\eIf{Continuous relaxation is integer feasible}{
Obtain the values of the binary variables $y_{ko}$ and continuous variables $c_{n}$\;
Construct the Lyapunov function $V(x)$\;
Find the maximum value of $\dot{V}(x)$, $v^{*} \in \arg \max_{x\in \mathcal{X}} \ \dot{V}(x) $\;
Find the minimum value of $V(x)$, $\hat{v}^{*} \in \arg \min_{x\in \mathcal{X}} \ V(x) $\;
Compute $V(0)$\;
\eIf{$v^{*} \leq 0$ \ \rm{and} $\hat{v}^{*} \geq 0$ \rm{and} $V(0)=0$}{
Stop search\;
Return Lyapunov function $V(x)$\;}
{Continue with branch and bound\;}
}
{Continue with branch and bound\;}

\eIf{$R= \varnothing$}{
\eIf{A valid Lyap. func. has been found}{Return $V(x)$\;}{Add more data and resolve the problem\;}
}{Continue\;}
}
\caption{Branch-and-bound-and-check algorithm for Lyapunov function discovery}
\label{algo: branch and bound and check}
\end{algorithm}

\begin{remark}
    \normalfont The performance of the proposed algorithm depends on the available data, the branching strategy of the solver, and the objective of the learning problem. Global optimization solvers use a set of heuristics for node selection, fathoming, and branching. These heuristics are designed such that the global optimal solution is found as fast as possible. However, as discussed above, the global optimal solution of the learning task will not necessarily provide a valid Lyapunov function. Therefore, even if the Lyapunov function is in the feasible space of the solver, it might not be found depending on the branching strategy of the solver. In such cases, one can add integer cuts to exclude certain functional forms for the Lyapunov function and resolve the problem. 
\end{remark}

\section{Computational results} \label{sec: computational resulrs}
In this section, we use the proposed approach to discover Lyapunov functions for various dynamical systems and explore its computational efficiency across different numbers of data points and objectives. In all the cases presented below, the learning task is modeled using Pyomo \cite{bynum2021pyomo} and solved using Gurobi v. 11.0.0 \cite{gurobi} with \verb|callbacks| to verify the validity of a candidate function once an integer feasible solution is found. Moreover, we adjust the solver's branching priority by prioritizing the binary variables. The dynamical systems are simulated using the explicit Runge-Kutta method.

\begin{table*}[h!]
\centering
\caption{Discovered Lyapunov function for the dynamical system presented in Eq.~\ref{eq: case 1 ode} for different sampling times $(T=5)$}
\begin{tabular}{cccccccc}
\cline{1-7}
\begin{tabular}[c]{@{}c@{}} $N_{\rm{data}}$\\ \end{tabular} & \begin{tabular}[c]{@{}c@{}}Sampling time\\ $\Delta t = T/N_{\rm{data}}$\end{tabular} & \begin{tabular}[c]{@{}c@{}}Solution \\ time (sec)\end{tabular} & \begin{tabular}[c]{@{}c@{}}Discovered\\ function\end{tabular} & \begin{tabular}[c]{@{}c@{}}Number of \\ variables\end{tabular} & \begin{tabular}[c]{@{}l@{}}Number of \\ constraints\end{tabular}& \begin{tabular}[c]{@{}c@{}}Number of branch \\ and bound nodes \end{tabular} \\ \hline
5   & 1   & 0.1    &  $(3x_{1}+x_{2})(2x_{1}+x_{2})$ & 177 (66 bin.)      &   598 & 810  \\
10    & 0.5      & 0.23 & $x_{1}^{2}+4x_{2}^{2}$  & 257 (66 bin.)  &  1048 & 383    \\
20    & 0.25      & 2  & $x_{1}^{2}+4x_{2}^{2}$  & 417 (66 bin.)  & 1948 & 3193  \\
50    & 0.1      & 42  & $x_{1}^{2}+4x_{2}^{2}$  & 897 (66 bin.)  & 4648 & 33085   \\
100    & 0.05      & 206              & $x_{1}^{2}+4x_{2}^{2}$  & 1697 (66 bin.)  &  9148 & 48170  \\
\hline
\end{tabular} \label{table: effect of sampling time}
\end{table*}

\subsection{Case 1}
\begin{figure}
    \centering
    \includegraphics[scale=0.5]{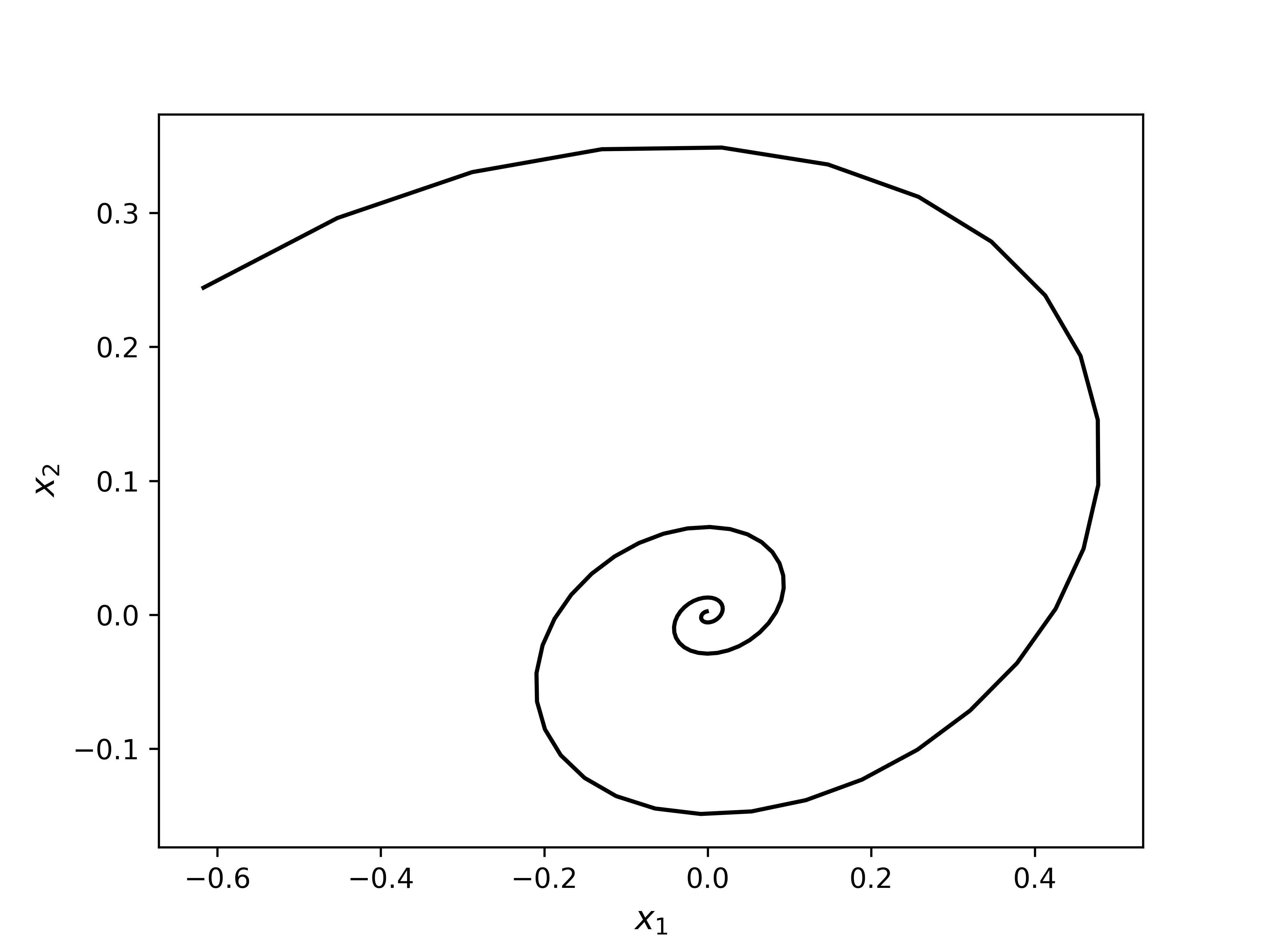}
    \caption{Learning trajectory for case 1}
    \label{fig: trajecotry case 1}
\end{figure}

First, we consider a dynamical system described by the following system of ordinary differential equations
\begin{equation}\label{eq: case 1 ode}
    \begin{aligned}
        \dot{x}_{1}(t) & = -x_{1}(t) + 4 x_{2}(t)\\
        \dot{x}_{2}(t) & = -x_{1}(t) - x_{2}(t)^{3}.
    \end{aligned}
\end{equation}
We generate one trajectory for $t \in [0,10]$ with 100 discretization points. The initial condition is sampled randomly between $[-2,2]$ for both state variables and is equal to $\textbf{x}_{0} = [-0.61 \ \   0.24]^\top$. The evolution of the states is presented in Fig.~\ref{fig: trajecotry case 1}.

The binary operators are $\mathcal{B}=\{+,-,*\}$, and the bounds for the variables are $c_{\rm{lb}}=1$, $c_{\rm{up}}=5$, $v_{\rm{lb}}=-50$, $v_{\rm{up}}=50$. We note that although the output of the Lyapunov function is nonnegative, the intermediate values can be negative depending on the assigned operators. Regarding the weights of the different terms in the objective of the learning task, we set $w_{1}=1$, $w_{2}=0$, and $w_{3}=0.1$, i.e., we minimize the weighted sum of the convergence rate and magnitude of the constants in the Lyapunov function. For this set of operators, the learning problem is a non-convex mixed integer quadratically constrained quadratic optimization problem (MIQCQP). 

First, we set the depth of the tree equal to 1 and the problem has 419 variables (12 binary) and 1828 constraints. Gurobi declares this problem as infeasible after presolve. Next, we increase the depth of the tree to two (seven nodes in the expression tree) and the problem is declared infeasible again after 1.5 seconds. These results show that a Lyapunov function with depth one or two, i.e., three or seven nodes in the expression tree, does not exist for the given data and approximation of the Lyapunov function time derivative. 

\begin{figure}[t]
    \centering
    \includegraphics[scale=0.6]{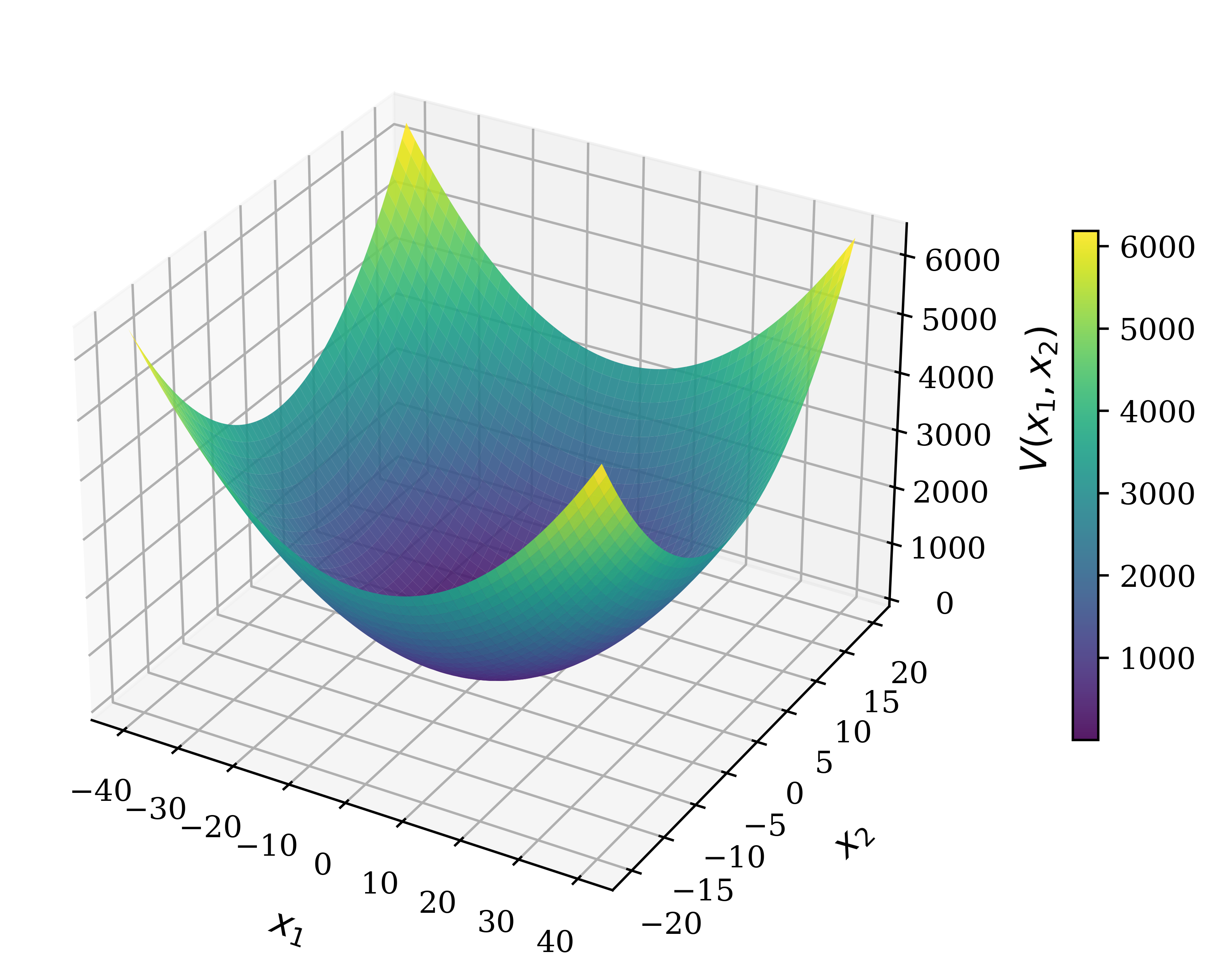}
    \caption{Plot of identified Lyapunov function $V(x_{1},x_{2}) = 2 x_{1}^{2}+8x_{2}^{2}$ for the dynamical system in case 1.}
    \label{fig: Lyapunov case 1}
\end{figure}

\begin{figure}[t]
    \centering
    \includegraphics[scale=0.6]{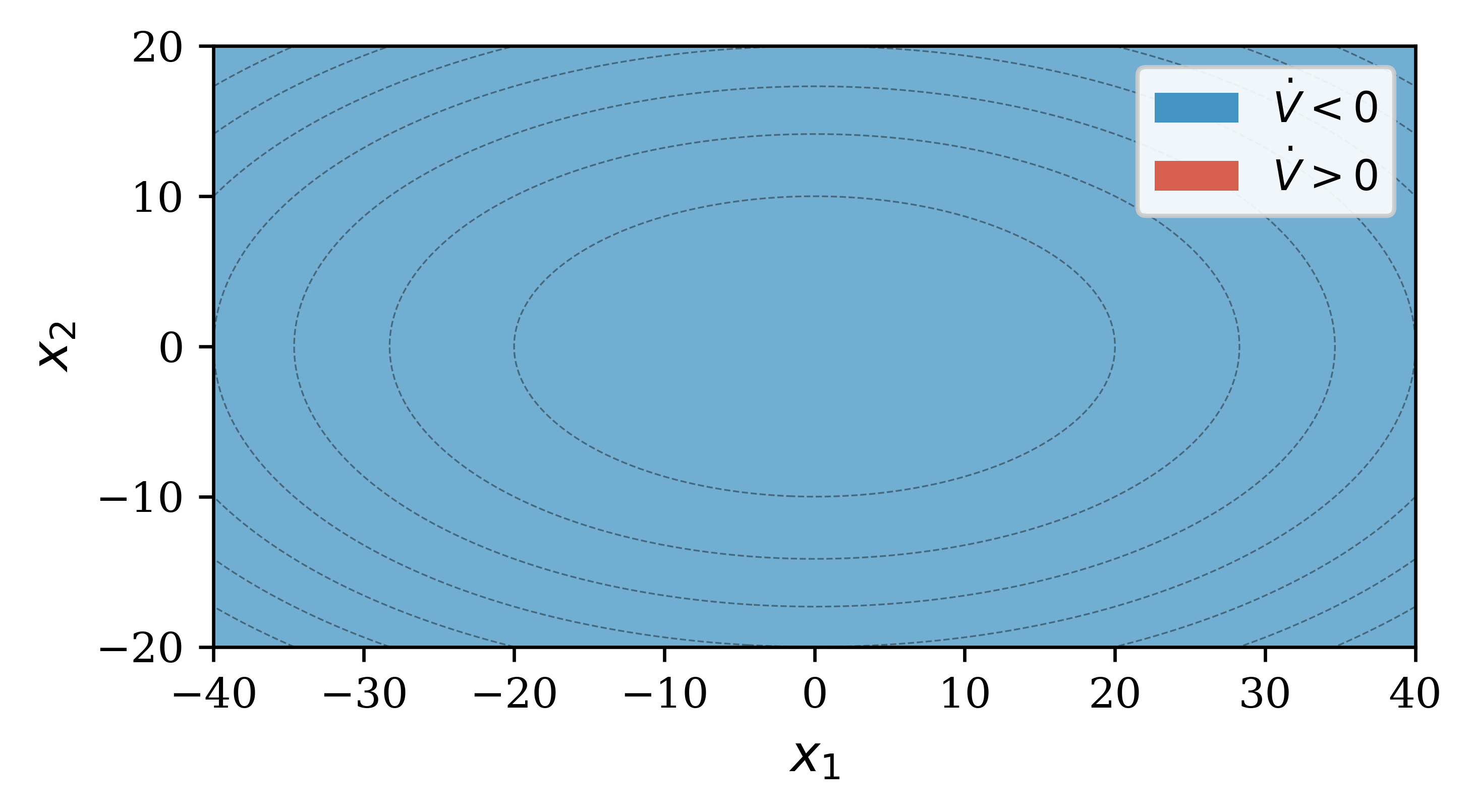}
    \caption{Sign of $dV/dt$ for the identified Lyapunov function $V(x_{1},x_{2}) = 2 x_{1}^{2}+8x_{2}^{2}$ for the dynamical system in case 1.}
    \label{fig: Lyapunov case 1 dvdt sign}
\end{figure}

We increase the depth of the tree to three (15 nodes) and the optimization problem has 1697 variables (66 binary) and 10562 constraints. After 45 seconds, the following candidate Lyapunov function is found
\begin{equation}
    V(x_{1},x_{2}) = (x_{1}+x_{1})x_{1} + (x_{2}+x_{2})4x_{2} = 2x_{1}^{2} + 8 x_{2}^{2},
\end{equation}
with $\alpha = 0.194$. This is a valid Lyapunov function for the entire domain by checking $V(0,0)=0$, $V(x_{1},x_{2}) >0 \ \forall x\neq 0$, and $dV/dt < 0 \ \forall x \neq 0$. The identified function and sign of $dV/dt$ are plotted in Fig.~\ref{fig: Lyapunov case 1} and \ref{fig: Lyapunov case 1 dvdt sign}.
If the search continues, a second valid Lyapunov function is found after 85 seconds 
\begin{equation}\label{eq: Lyapunov function for case 1 ode}
    V(x_{1},x_{2}) = (x_{2}+x_{2})(x_{2}+x_{2}) + x_{1}x_{1} = x_{1}^{2} + 4 x_{2}^{2}
\end{equation}
with $\alpha = 0.10$. 

Next, we analyze the effect of the weights on the solution of the problem. We fix the convergence rate to $\alpha=0.1$ and set $w_{3}=1$ and $w_{1} = w_{2}=0$. The problem is solved to global optimality in 40 seconds, and the following Lyapunov function is returned
\begin{equation}
    V(x_{1},x_{2}) = x_{1}*(x_{1}*1.0)+((x_{2}+x_{2})*(x_{2}+x_{2})) = x_{1}^{2}+4x_{2}^{2},
\end{equation}
which is a valid Lyapunov function for the entire domain.

\subsection{Effect of sampling rate}
We analyze the effect of the sampling rate on the solution of the learning task. As discussed in Section~\ref{sec: lyapunov defining constraints}, the time derivative of the Lyapunov function is approximated with finite differences. To evaluate the effect of this modeling approach on the solution of the learning task, we fix the time horizon length to five time units and vary the number of discretization points. For this analysis, we fix $\alpha=0.1$ and set $w_{1}=w_{2}=0$ and $w_{3}=1$. The results are presented in Table~\ref{table: effect of sampling time} and all the problems are solved to global optimality, i.e., the search does not stop once an integer feasible solution is found. From the results, we observe that as the number of data points increases, i.e., sampling time decreases, the solution time increases due to the increase in the number of variables and constraints. We note that the number of binary variables remains constant, since it depends on the depth of the tree and the operators. For a small number of data points, $N_{\rm{data}}=5$ the problem is solved in 0.1 seconds, but the returned function is not a valid Lyapunov function for the entire domain. In this case, the discretization step is 1 time unit making the approximation of the time derivative of the Lyapunov function with finite differences inaccurate. As the number of data points increases to $N_{\rm{data}}=20$, the returned function is a valid Lyapunov function for the entire domain. These results show that an increase in the sampling rate aids the discovery of the Lyapunov function, as the approximation of its time derivative becomes more accurate. Finally, we note that the number of nodes in the branch-and-bound tree explored increases with the number of data points. Although we alter the solver's branching priorities, it is not guaranteed that Gurobi strictly follows them; the solver might branch on the continuous variables that capture the intermediate values at each node.

\subsection{Effect of the time horizon length}
We also analyze the effect of the time horizon length on discovering a Lyapunov function. Specifically, we fix the discretization step to $\Delta t= 0.1$ and generate a single trajectory for $T$ time units and the results are presented in Table~\ref{table: effect of time horizon}. From the results we observe that for short time horizons, the returned function is not a valid Lyapunov function for the entire domain. Specifically, the functions returned for $T=1$, $2$, and $3$ are not valid Lyapunov functions for the entire domain. These functions correspond to the globally optimal solution of each learning task and during the branch-and-bound-and-check search, a Lyapunov function that is valid for the entire domain is not found. However, if the length of the time horizon is increased to five, then a valid Lyapunov function is discovered. These results show the importance of capturing the underlying dynamic behavior of the system, since in all cases the discretization step was fixed at $0.1$. 

\begin{table*}[h]
\centering
\caption{Discovered Lyapunov function for the dynamical system presented in Eq.~\ref{eq: case 1 ode} for different time horizon lengths. The discretization step is $0.1$.}
\begin{tabular}{cccccccc}
\cline{1-7}
\begin{tabular}[c]{@{}c@{}} $T$\\ \end{tabular} & \begin{tabular}[c]{@{}c@{}}$ N_{\rm{data}}$\end{tabular} & \begin{tabular}[c]{@{}c@{}}Solution \\ time (sec)\end{tabular} & \begin{tabular}[c]{@{}c@{}}Discovered\\ function\end{tabular} & \begin{tabular}[c]{@{}c@{}}Number of \\ variables\end{tabular} & \begin{tabular}[c]{@{}l@{}}Number of \\ constraints\end{tabular}& \begin{tabular}[c]{@{}c@{}}Number of branch \\ and bound nodes \end{tabular} \\ \hline
1   &  10  & 0.2    &  $x_{1}^{2}+2x_{2}$ & 257 (66 bin.)      &   1048 & 264  \\
2    & 20      & 3 & $x_{1}^{2}+3x_{2}^{2}$  & 417 (66 bin.)  &  1948 & 6562    \\
4    & 40      & 178  & $x_{1}^{2}+4.12x_{2}^{2}$  & 737 (66 bin.)  & 3748 & 185062  \\
5    & 50      & 42  & $x_{1}^{2}+4x_{2}^{2}$  & 897 (66 bin.)  & 4648 & 33085   \\
\hline
\end{tabular} \label{table: effect of time horizon}
\end{table*}

\subsection{Effect of the number of initial conditions}
\begin{figure}
    \centering
    \includegraphics[scale=0.5]{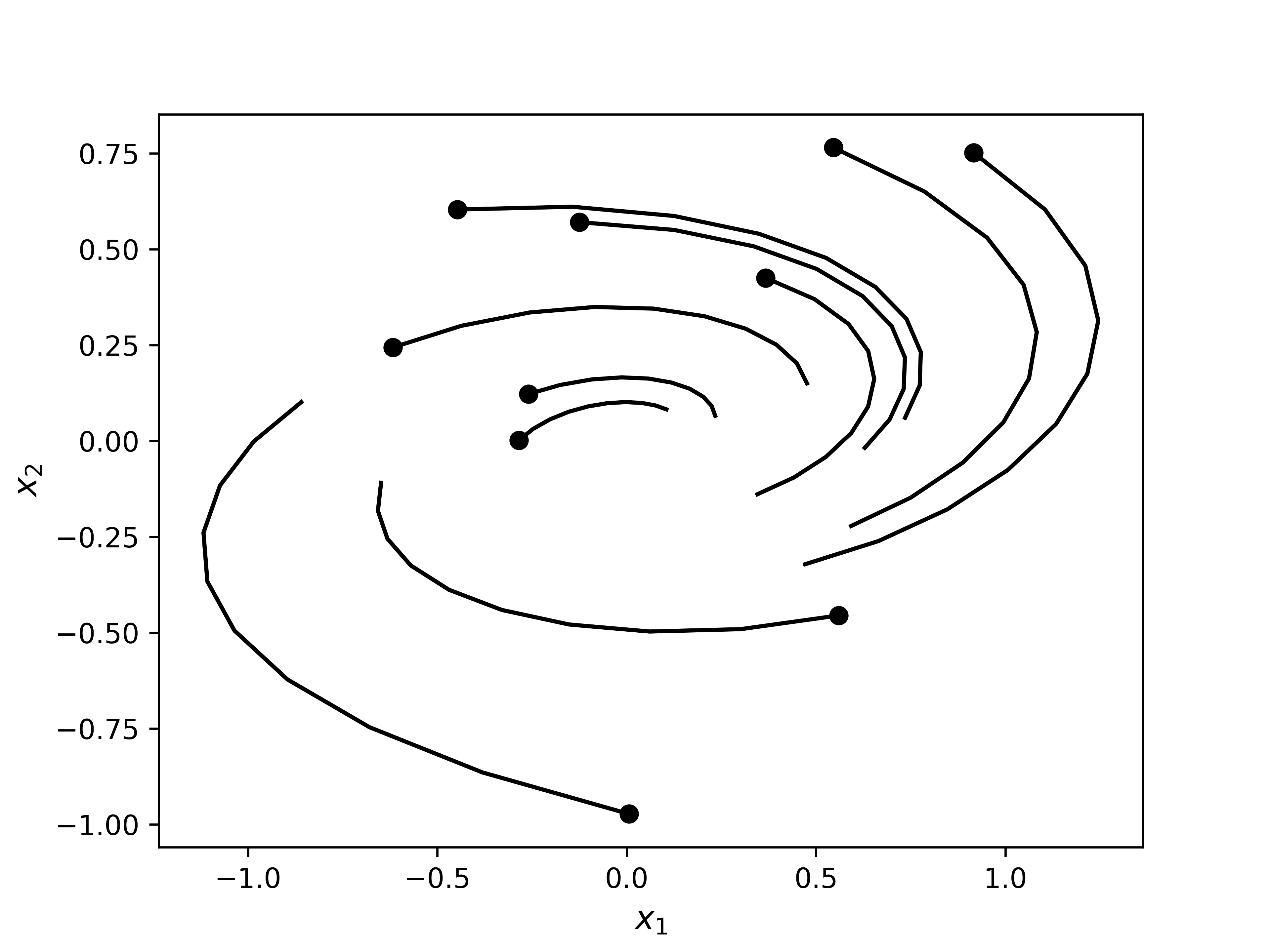}
    \caption{Learning trajectory for case 1 using snapshots from 10 initial conditions with time horizon equal to 1 and discretization step equal to 0.1. The black dots correspond to the initial conditions.}
    \label{fig: snapshot trajectory case 1 snapshots}
\end{figure}

In the results presented so far, the training data included a single trajectory. In this section, we analyze how the number of initial conditions and time horizon length affect the ability of the proposed approach to discover a Lyapunov function. For all cases, we fix the convergence rate to $\alpha=0.1$ and set $w_{1}=0$, $w_{2}=0$, and $w_{3}=1$. We sample uniformly the initial conditions between $[-1, 1]$ for both state variables (see Fig.~\ref{fig: snapshot trajectory case 1 snapshots} as an example). All the problems are solved to global optimality and the results are presented in Table~\ref{table: effect of number of initial conditions}. 

From the results, first we observe that as the number of initial conditions, $N_{\rm{t}}$, increases the solution time increases since the number of continuous variables and associated value and Lyapunov defining constraints increases. For the case where the time horizon is $0.2$ time units, we observe that for five initial conditions, the returned function is not valid for the entire domain. Increasing the number of initial conditions to 10 leads to a valid Lyapunov function. 

Finally, we increase the length of the time horizon to one time unit and solve the problem for different number of initial conditions. We observe that for three initial conditions, the returned function is not a valid Lyapunov function. This result shows that although the duration of each trajectory is longer than the case with $T=0.2$, more trajectories are needed. Increasing the number of trajectories to five yields a valid Lyapunov function.

\begin{table}[h]
\centering
\caption{Discovered Lyapunov function for the dynamical system presented in Eq.~\ref{eq: case 1 ode} for different number of initial conditions $(N_{\rm{t}})$ and time horizon length $(T)$. The discretization step is $0.1$.}
\begin{tabular}{cccc}
\cline{1-4}
\begin{tabular}[c]{@{}c@{}} $T$\\ \end{tabular} & \begin{tabular}[c]{@{}c@{}}$ N_{\rm{t}}$\end{tabular} & \begin{tabular}[c]{@{}c@{}}Solution \\ time (sec)\end{tabular} & \begin{tabular}[c]{@{}c@{}}Discovered\\ function\end{tabular}  \\ \hline
0.2   &  5  & 0.3    &  $x_{1}^{2}+2x_{2}$ \\
0.2   &  10  & 2    &  $x_{1}^{2}+4x_{2}^{2}$ \\
0.2   &  15  & 8    &  $x_{1}^{2}+4x_{2}^{2}$ \\
1    & 3      & 2 & $x_{1}^{2}+3x_{2}^{2}$  \\
1    & 5      & 3.5  & $x_{1}^{2}+4x_{2}^{2}$  \\
1    & 10      & 9  & $x_{1}^{2}+4x_{2}^{2}$   \\
1    & 15      & 152  & $x_{1}^{2}+4x_{2}^{2}$   \\
\hline
\end{tabular} \label{table: effect of number of initial conditions}
\end{table}

\section{Discovering the simplest Lyapunov function}
In this section, we use the proposed approach to search for the simplest Lyapunov function, i.e., the function with the fewest number of operators in the expression tree. 

\subsection{Minimizing the summation of the binary variables}
An approach to identify the simplest Lyapunov function is to minimize the summation of the binary variables as follows
\begin{equation} \label{eq: learning task simple}
    \begin{aligned}
    \min_{y_{ko}, c_{n}, v_{ijn}, 
    \alpha} \ \ & \sum_{n \in \mathcal{N}} \sum_{o \in \mathcal{O}} y_{no}\\
    \text{s.t.} \ \ & \mathrm{Eq.}~\ref{eq: one operator}-\ref{eq: neg def V}\\
    & y_{no}\in\{0,1\}, v_{ijn} \in [v_{lo}, v_{up}]\\
    & c_{n} \in [c_{lo}, c_{up}], \alpha \geq 0.
    \end{aligned}
\end{equation}
We use the dynamical system from Eq.~\ref{eq: case 1 ode} and generate a single training trajectory simulating the system for $t \in [0,10]$ using 100 discretization points and fixing $\alpha=0.1$.

The first integer feasible solution is found after 130 seconds and corresponds to the following Lyapunov function
\begin{equation}
    V(x_{1},x_{2}) = (x_{2}*3)*x_{2}+ x_{2}*x_{2}+x_{1}*x_{1} = x_{1}^{2}+4x_{2}^{2},
\end{equation}
which is a valid Lyapunov function with 13 nodes in the expression tree. After 227 seconds the following function is found with nine nodes in the expression tree
\begin{equation}
    V(x_{1},x_{2}) =  ((x_{2}*3.97)*x_{2})+(x_{1}*x_{1}) = x_{1}^{2}+3.97x_{2}^{2}.
\end{equation}
This function, although has the right functional form the coefficient of the $x_{2}^{2}$ term is not correct. Although this difference can be attributed to limited data and the approximation of the derivative of the Lyapunov function, another possible issue is the degeneracy of the learning task. Specifically, when the objective is the sum of the binary variables, the values of the constants in the expression tree do not affect the objective function value at the optimal solution. Therefore, multiple parameter values can yield the same objective as long as the value and Lyapunov-defined constraints are satisfied.

\subsection{Constraining the complexity of the Lyapunov function}
An alternative to minimizing the summation of the binary variables is to constrain the sum of the binaries as follows
\begin{equation}
    \sum_{n \in \mathcal{N}} \sum_{o \in \mathcal{O}} y_{no} \leq N_{\rm{C}},
\end{equation}
with $N_{\rm{C}}$ being the maximum number of operators in the expression tree, i.e., the maximum number of operations that must be performed to compute the output of the tree. The resulting learning task is
\begin{equation} \label{eq: learning task simplest V}
    \begin{aligned}
    \min_{y_{ko}, c_{n}, v_{ijn}, 
    \alpha} \ \ & -\alpha \\
    \text{s.t.} \ \ & \mathrm{Eq.}~\ref{eq: one operator}, \ref{eq: operator assignment 1}, \ref{eq: operator assignment 2}, \ref{eq: assign x variables}, \ref{eq: value def addition}, \ref{eq: value defining cst}, \ref{eq: value defining sub}, \ref{eq: value defining mult}, \ref{eq: value defining division}, \ref{eq: value defining x 1}, \ref{eq: value defining x 2}, \ref{eq: K class function}, \ref{eq: neg def V}\\
    & \sum_{n \in \mathcal{N}} \sum_{o \in \mathcal{O}} y_{no} \leq N_{\rm{C}}\\
    & y_{no}\in\{0,1\}, v_{ijn} \in [v_{lo}, v_{up}]\\
    & c_{n} \in [c_{lo}, c_{up}], \alpha \geq 0,
    \end{aligned}
\end{equation}
We generate $N_{\rm{t}}=15$ training trajectories for $t \in [0,1]$ using 20 discretization points and solve the resulting learning task for different maximum number of operators in the expression tree. The depth of the tree is fixed to three, i.e., the maximum number of nodes and thus operators in the expression is 15. From Table~\ref{table: search for simplest tree} we observe that for all cases, the discovered function is a valid Lyapunov function for the entire domain. For $N_{\rm{C}}=15$ the identified function is $2x_{1}^{2}+8x_{2}^{2}$ which is the same as the one identified in Section~4.1. For $N_{\rm{C}}=13$ the identified function is $x_{1}^{2}+4x_{2}^{2}$ whereas for $N_{\rm{C}}=11$ the function is $1.25x_{1}^{2}+5x_{2}^{2}$. Comparing the solution time, we observe that as the maximum number of operators decreases, the solution time also decreases since the search space is more constrained. We note that if the tree depth is 2, which corresponds to a maximum of 7 nodes, the learning task is infeasible. Therefore, the simplest Lyapunov function has nine nodes for the given set of operators, i.e., a tree with operators at eight nodes is not feasible since we only consider binary operators and operands.
\begin{table}[h]
\centering
\caption{Discovered Lyapunov function for the dynamical system presented in Eq.~\ref{eq: case 1 ode} for different maximum number of operators in the expression tree. }
\begin{tabular}{ccc}
\cline{1-3}
\begin{tabular}[c]{@{}c@{}} $N_{\rm{C}}$\\ \end{tabular}& \begin{tabular}[c]{@{}c@{}}Solution \\ time (sec)\end{tabular} & \begin{tabular}[c]{@{}c@{}}Discovered\\ function\end{tabular}  \\ \hline
15   &  74    &  $(x_{1}*x_{1})+(x_{1}*x_{1})+(x_{2}+x_{2})*(4*x_{2})$ \\
13   &   65    & $5*(x_{2}*x_{2})+(x_{1}*x_{1})-(x_{2}*x_{2})$ \\
11   &   61    &  $1.25*(x_{1}*x_{1})+(x_{2}*5)*x_{2}$ \\
9    &  20 & $x_{1}*x_{1}+(x_{2}*4)*x_{2}$  \\ \hline
\end{tabular} \label{table: search for simplest tree}
\end{table}

\section{Lyapunov function discovery as a feasibility problem}
\begin{figure}[t]
    \centering
    \includegraphics[scale=0.5]{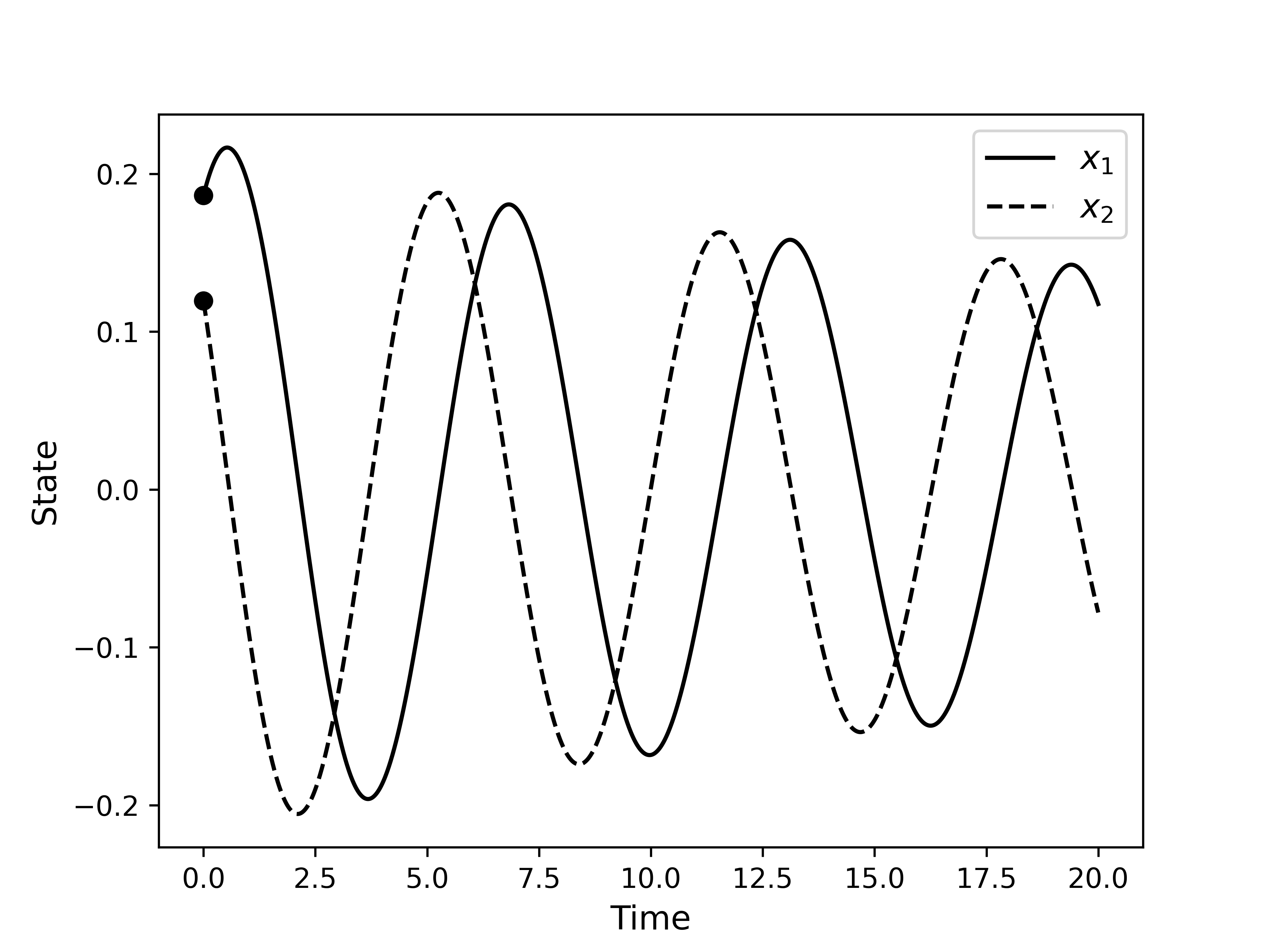}
    \caption{State trajectories for case 2.}
    \label{fig: snapshot trajectory case 2}
\end{figure}

\begin{figure}[t]
    \centering
    \includegraphics[scale=0.5]{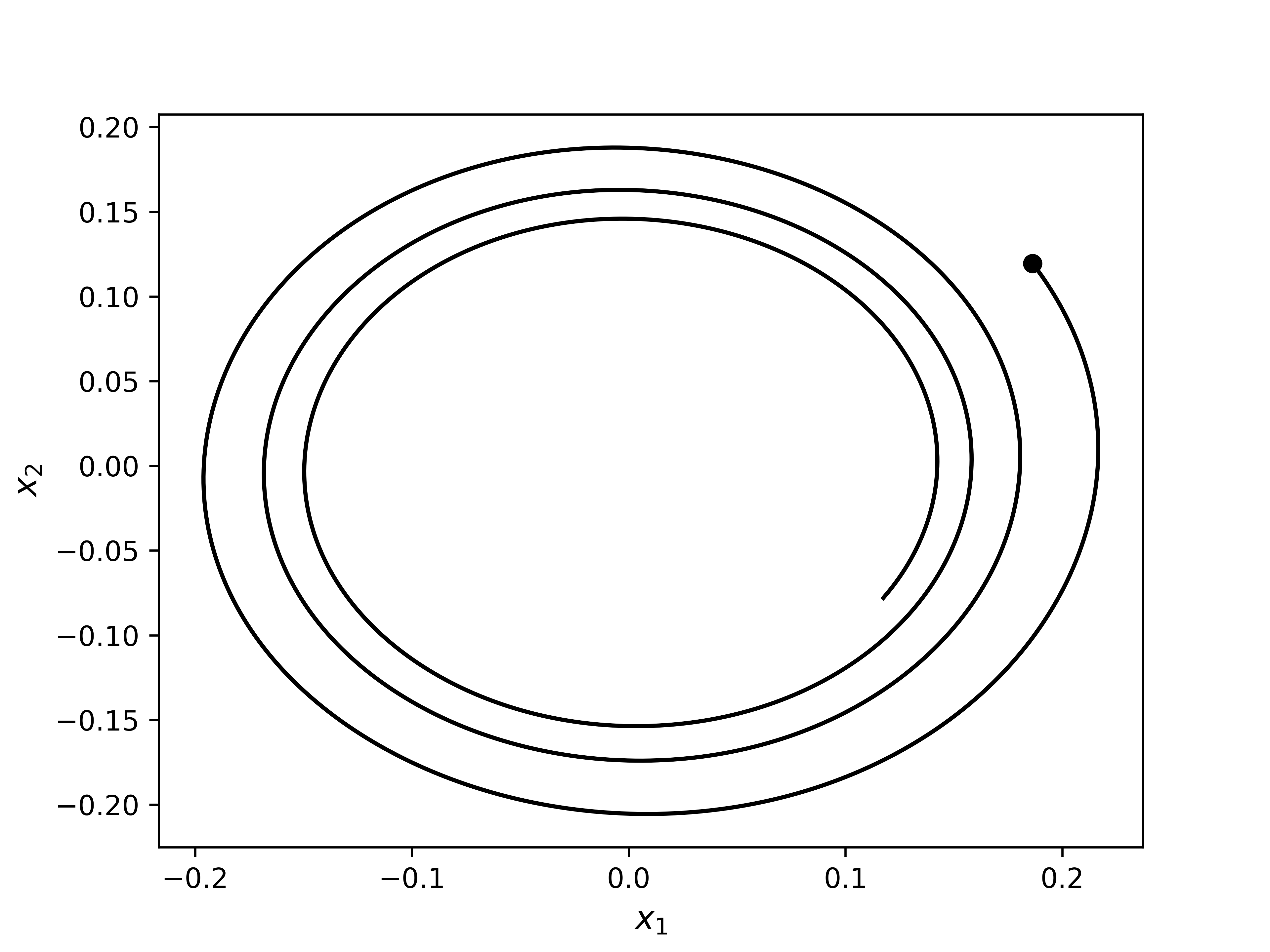}
    \caption{Phase portrait for the case 2 dynamical system.}
    \label{fig: snapshot phase plane case 2}
\end{figure}

Finally, we analyze the performance of the proposed approach if the learning task is posed as a feasibility task. We consider the following dynamical system (case 2)
\begin{equation}
    \begin{aligned}
        \dot{x}_{1}(t) & = x_{2}(t) - x_{1}(t)^{3}\\
        \dot{x}_{2}(t) & = -x_{1}(t) - x_{2}(t)^{3}.
    \end{aligned}
\end{equation}
The evolution of the state variables from the initial condition of $\textbf{x}_{0} = [0.18 \ \  0.11]^\top$ is presented in Fig.~\ref{fig: snapshot trajectory case 2} and Fig.~\ref{fig: snapshot phase plane case 2}.

For the learning task, we generate one trajectory for $t \in [0,50]$ using 200 discretization points. First, we minimize the summation of the convergence rate and magnitude of Lyapunov function coefficients, $-\alpha + 0.1 \sum_{n \in \mathcal{N}} c_{n}^{2}$. The learning task has 3297 variables (66 binary) and 18148 constraints. The first integer feasible solution found during branch and bound after 8 seconds (93 nodes in the branch and bound tree) is
\begin{equation}
    V(x_{1},x_{2}) = x_{1}^{2}+x_{2}^{2},
\end{equation}
which is a valid Lyapunov function for the system with $\alpha = 0.0097$. 

Next, we set the objective of the learning task to $0$ and solve the resulting feasibility problem without using the above solution as a warm start. For the given data set, the problem is solved in 30 seconds (2272 nodes in the branch and bound tree), and the following Lyapunov function is found
\begin{equation}
    V(x_{1},x_{2}) = ((1.0*x_{1})*x_{1})+((x_{2}*1.0)*x_{2}))=x_{1}^{2}+x_{2}^{2},
\end{equation}
which is the same as the one obtained above; however, the solution time is higher.

\section{Conclusions}
This paper considers the data-driven discovery of Lyapunov functions. Specifically, we use the symbolic tree representation of the Lyapunov function to build a superstructure of candidate functions. We incorporate constraints that guarantee the correctness of the expression with respect to arity rules and the computation of the function's output for a given input and assignment of mathematical operators. Moreover, we incorporate constraints that enforce the Lyapunov stability conditions for the given data. Finally, we develop a branch-and-bound-and-check solution approach to efficiently solve the resulting problem. This solution strategy enables simultaneous search for a candidate Lyapunov function via branch-and-bound and verification by checking its validity across the entire domain. The ability of the proposed approach to discover Lyapunov functions is demonstrated in several case studies. 

The computational efficiency of the proposed approach depends on the number of state variables, the complexity of the Lyapunov function in terms of operators and depth of the expression tree, and the number of data points required to capture adequate information for the dynamic evolution of the system. Several approaches can be pursued to improve the tractability of the proposed approach. For example, one can leverage decomposition-based optimization algorithms, such as Benders decomposition \cite{geoffrion1972generalized, li2011nonconvex}, and use active learning-like approaches to adaptively incorporate data.

\section*{Acknowledgements}

IM would like to acknowledge financial support from the McKetta Department of Chemical Engineering at The University of Texas at Austin.

\bibliographystyle{elsarticle-num}
\bibliography{references}

\end{document}